\documentclass[12pt]{article}
\usepackage[pdftex]{graphicx}
\usepackage{amsmath}
\usepackage{xcolor,soul}
\usepackage{url}
\usepackage{amssymb,esint}

\title{\bf Photoelectric effect induced by blackbody radiation: a theoretical
  analysis of a potential heat energy harvesting mechanism}
\author{Germano D'Abramo\\
{\small Ministero dell'Istruzione, dell'Universit\`a e della Ricerca,}\\
{\small 00041, Albano Laziale, RM, Italy}\\
{\small E--mail: {\tt germano.dabramo@gmail.com}}}

\date{ORCID: 0000-0003-1277-7418}
  
\begin{document}

\maketitle

\begin{abstract}
  The photoelectric effect induced by blackbody radiation could be a
  mechanism to harvest ambient thermal energy at a uniform temperature. Here, I
  describe (without going too much into mathematical details) the theoretical
  model I developed starting from 2010 to study that phenomenon, and I
  summarize the results of the numerical simulations. Simulations tell us that
  the process must be there. Moreover, at least two experimental tests have
  been performed in the past years that seem to corroborate, although not
  definitely, the alleged functioning of the proposed mechanism. Unfortunately,
  at present, the obtainable power density is extremely low and inadequate for
  immediate practical applications.\\


\end{abstract}

\section{Introduction}
\label{se1}

If we can get energy from ambient light via the photoelectric effect, why should
it not be possible to harvest ambient thermal energy at a uniform temperature
through the photoelectric effect caused by blackbody radiation? In the
paper, I sometimes refer to this second process as the thermionic
emission\footnote{Thermionic emission is usually intended as the boiling off of
  electrons from a metal surface due to thermal energy (heat). It is a distinct
  phenomenon from the ejection of electrons due to the absorption of
  electromagnetic radiation, which is the photoelectric effect studied in the
  present paper. However, when one deals with blackbody radiation, the two
  phenomena are strictly intertwined. In fact, blackbody radiation is thermal
  electromagnetic radiation in thermodynamic equilibrium with the surrounding
  matter. Moreover, thermionic emission acts upon electrons with a similar
  mechanism (for emission, the thermal energy must be greater than the work
  function of the material) and works in the same direction as the effect
  studied in the present paper (both tend to eject electrons).}. What I will
present here is a theoretical analysis of that possibility, and it
stands on my work on that topic dating back to 2010~\cite{dab10,dab11a, dab11b,
dab12a,dab12b,dab13a,dab13b,dab17,leff}. I anticipate
that the answer is affirmative in principle. Unfortunately, the obtainable power
density  turns out to be extremely low and inadequate for immediate practical
applications. Due to its simplicity, this idea is not new. It has already been
proposed and even experimentally approached (Section~\ref{se5}), but usually
with what appears to be a not fully satisfying theoretical explanation. Here, I
describe my contribution to remedying this lack, also because, as
Sir.~Eddington put it, you should not put much confidence in experimental
results until they have been confirmed by theory\footnote{Obviously, this is a
joke. But, in some circumstances, it has an element of truth.}.

\section{The basic idea}
\label{se2}

\begin{figure}[t]
\begin{center}
\includegraphics[width=8cm]{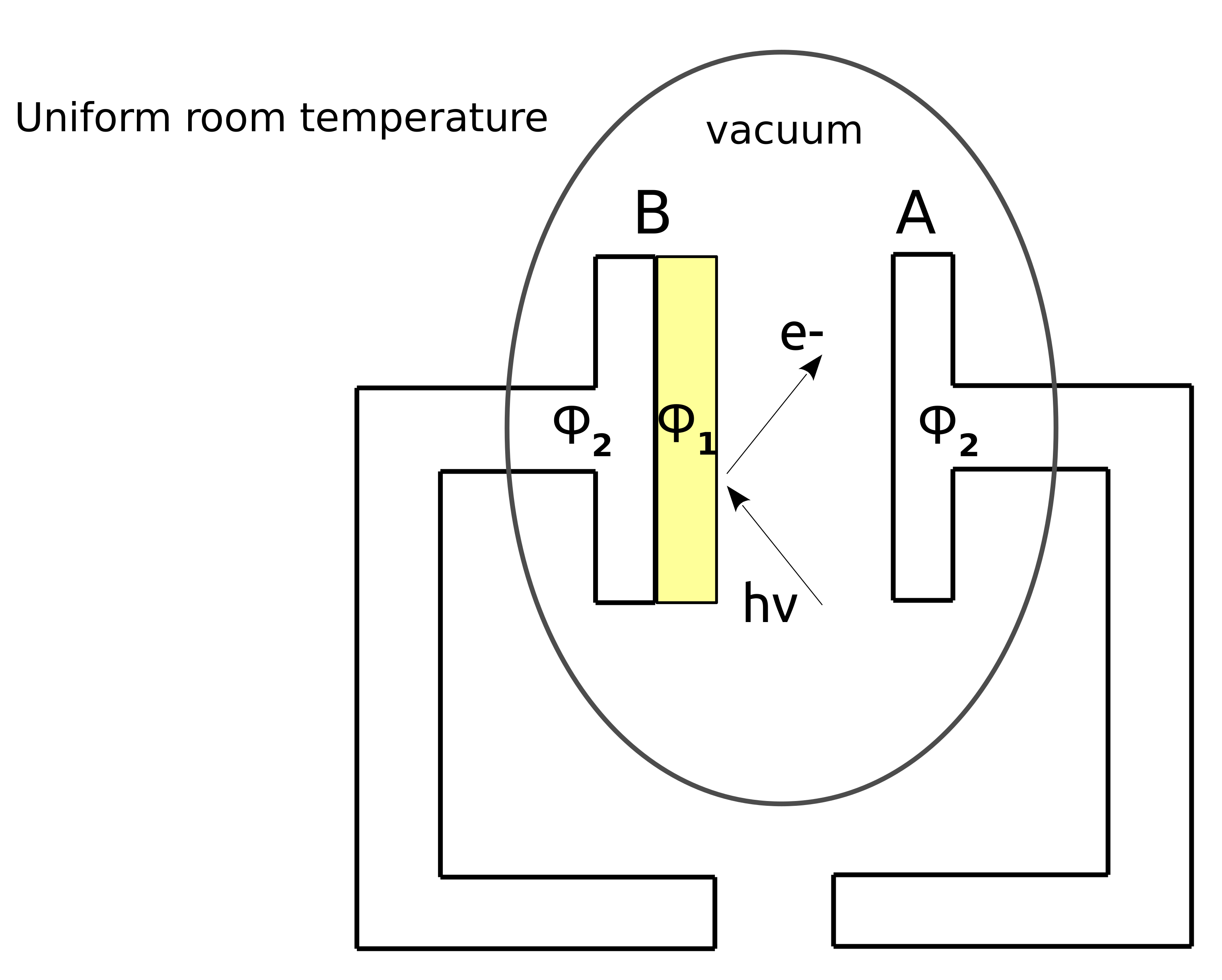}
\end{center}
\caption{Sketch of a thermo-charged capacitor (TCC)}
\label{fig1}
\end{figure}

The idea behind the approach proposed in this paper is simple: exploiting the
photoelectric effect of the ambient blackbody radiation (at a uniform
temperature) on materials with different work functions $\phi$\footnote{The
  work function $\phi$ is roughly the minimum energy (in eV) required to extract
  an electron from the bulk of a material to its surface.}. The simplest design
to do that is shown in Fig.~\ref{fig1}.

The sketched device is a capacitor, which I dubbed ``thermo-charged capacitor''
(or TCC), where two plates, A and B, are housed inside a vacuum bulb. Plate A
and part of plate B are made of the same metal with work function $\phi_2$.
Plate B is coated with a semiconductor with a lower work function $\phi_1$.
The whole device is at a uniform temperature.

Both plates A and B receive photons from the blackbody radiation, but being
$\phi_2$ greater than $\phi_1$, the electrons extracted by the photoelectric
effect from the coating of B to plate A are more numerous and with higher
kinetic energy $K$ ($K = h\nu - \phi$, where $h\nu$ is the photon energy) than
those extracted from plate A to plate B, at least in the early phases of
the process.

\begin{figure}[t]
\begin{center}
\includegraphics[width=8cm]{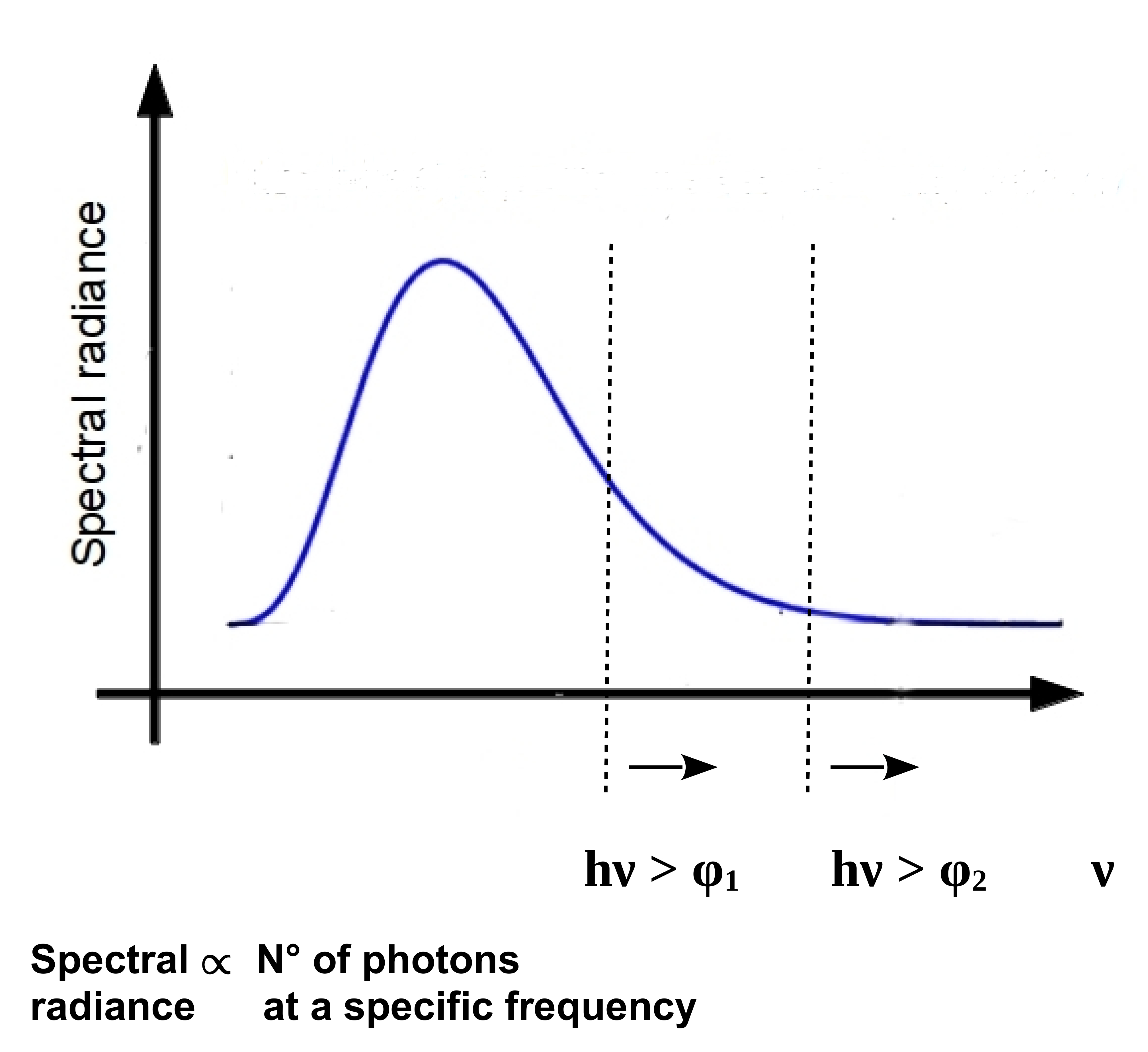}
\end{center}
\caption{Blackbody radiation at a uniform temperature. The spectral radiance is
  proportional to the number of blackbody photons at a specific frequency}
\label{fig2}
\end{figure}

This behavior is evident when we look at the shape of the spectral radiance
vs.~the frequency of the blackbody radiation at equilibrium (uniform
temperature), Fig.~\ref{fig2}.

The spectral radiance being proportional to the number of blackbody photons at
a specific frequency, the number of photons in the thermal radiation with
energy $h\nu > \phi_1$ is greater than the number of photons with
$h\nu > \phi_2$, and thus it is easy to see why more electrons are extracted
from B to A and with higher kinetic energy. All that means that an {\em
  unbalanced net flux} of electrons should flow from plate B to plate A. 

\begin{figure}[h]
  \includegraphics[width=7cm]{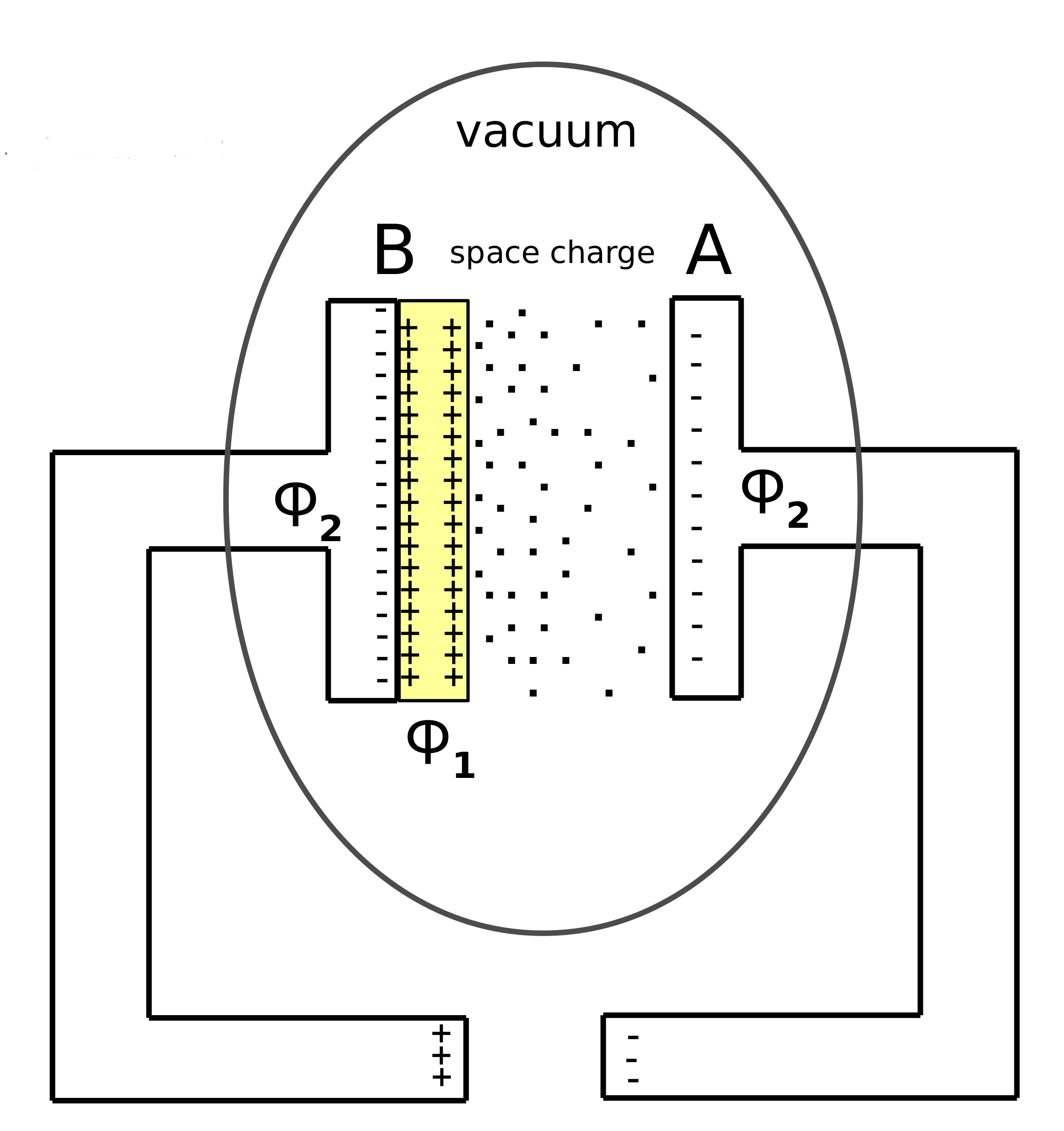}\hspace{0.5cm}
  \includegraphics[width=7cm]{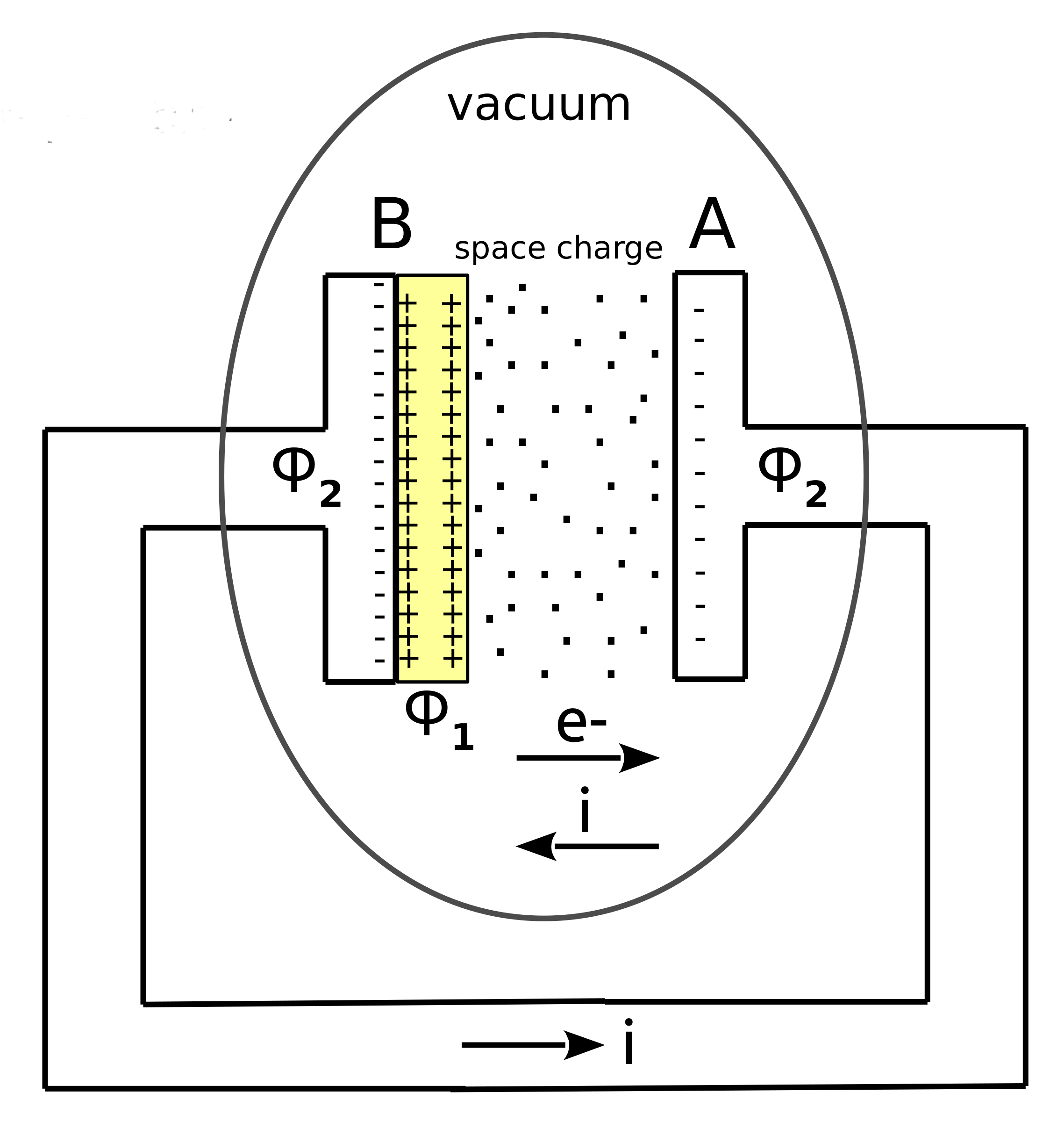}
\caption{Open-circuit (left) and short-circuit (right) TCCs}
\label{fig3}
\end{figure}

To show how the TCC is expected to work, it may help to consider the following
two distinct setups: open-circuit and short-circuit TCC, see Fig.~\ref{fig3}.
In the case of an open-circuit TCC (Fig.~\ref{fig3}, left), we expect that
after a suitable interval of time, an equilibrium potential difference
$\Delta V$ gets established across the plates.

The equilibrium potential difference $\Delta V$ is necessarily equal to
$(\phi_2 - \phi_1)/e$, where $e$ is the electron charge, because only with that
value the energy $E_B$ needed by an electron inside the coating to get
extracted and reach plate A ($E_B = \phi_1 + e\Delta V = \phi_1 + e(\phi_2 -
\phi_1)/e = \phi_2$) becomes equal to the energy $E_A$ needed by an electron to
get extracted from plate A and reach plate B (only $\phi_2$). Only in this case,
there will be no more imbalance in the two fluxes of electrons.

In the case of a short-circuit TCC (Fig.~\ref{fig3}, right), a current $i$
should flow in the circuit. If the resistance of the circuit external to the
TCC is negligible, that current is equal to the thermionic current
$i_{thermionic}$ across the plates, and the potential difference between the
plates is nearly equal to  zero.

The small dots between the plates represented in Fig.~\ref{fig3}, especially in
the open-circuit TCC configuration, is the space charge. In stationary
conditions like that of a fully charged open-circuit TCC, electrons are
continuously emitted and reabsorbed by the plates in a sort of dynamical
equilibrium, and thus they form a cloud. This is well-known among those who used
to work with vacuum tubes. Space charge distribution inside a TCC can be modeled
mathematically, but it is not difficult to understand that in an open-circuit
configuration at equilibrium, the electron cloud density decreases from plate
B to plate A (fewer and fewer electrons in plate B get enough energy from the
blackbody radiation to get closer and closer to plate A). 

Notice that this space charge does not change the value of the total stationary
potential drop between the plates, which always remains equal to
$(\phi_2 - \phi_1)/e$ for the reasons explained above. The space charge
interferes only with the uniformity of that potential in the inter-plate space:
the potential difference does not change linearly with the distance from the
plates as it happens in a standard charged parallel-plate capacitor. 

\section{Possible objections to the expected functioning}
\label{se3}

Before presenting the results of the mathematical and numerical modeling of
both open-circuit and short-circuit TCCs, let me dwell on two main objections
that could be advanced against the functioning described above. It is
fundamental to face the following objections and reply to them because, if they
were correct, the TCC would not work as expected. 

\begin{figure}[t]
\begin{center}
  \includegraphics[width=6cm]{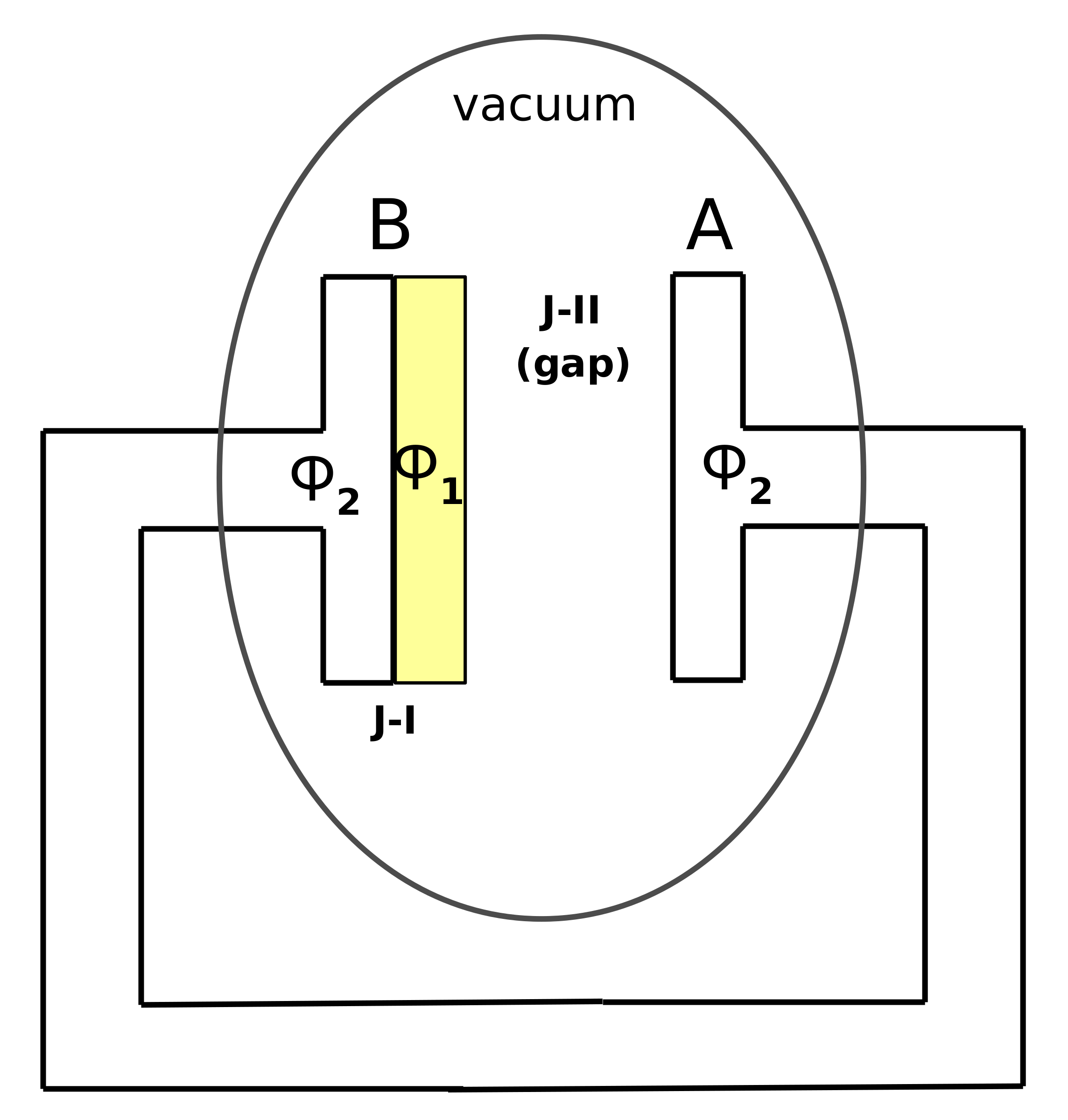}\hspace{1cm}
  \includegraphics[width=6cm]{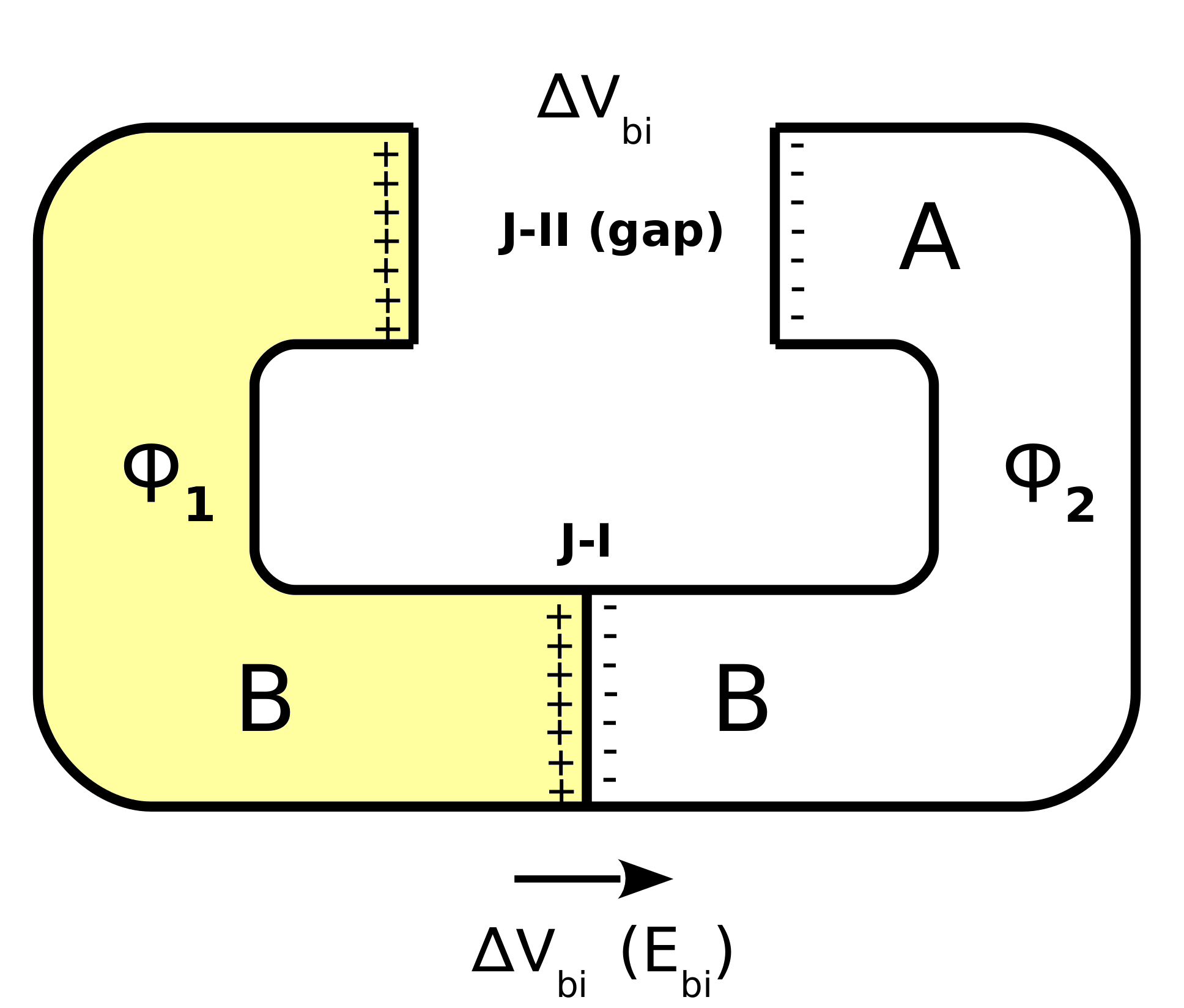}
\end{center}
\caption{Short-circuit TCC (left) and its topological analog (right)}
\label{fig4}
\end{figure}

To better introduce them, I will use a topological analog of the short-circuit
TCC in Fig.~\ref{fig3} (right). I refer to the horseshoe-shaped scheme
shown on the right-hand side of Fig.~\ref{fig4}. The right half of the
horseshoe represents plate A, part of plate B, and the connecting wire, and
all this stuff has a work function equal to $\phi_2$. The left half is the
coating of plate B with work function $\phi_1$. In both images, J-II indicates
the vacuum gap between the plates. 

It is known that when two materials with different work functions are
physically joined, a potential difference $\Delta V_{bi}$ builds up across the
contact junction J-I (Fig.~\ref{fig4}, right). This potential is known as the
{\em contact potential}, and it is generated by electrons that are pushed from
the $\phi_1$ material to the $\phi_2$ material across J-I by thermally driven
forces (thermal agitation). 

A higher amount of electrons move from left to right rather than the other way
around simply because it’s easier to pull electrons out of a lower work function
material. This process lasts until a dynamical equilibrium is reached between
the thermally driven forces and the electric force due to the built-in
electric field across J-I. 

Also, in this case, the contact potential $\Delta V_{bi}$ across J-I is equal to
$(\phi_2 - \phi_1)/e$, and the reason is the same as that shown in
Section~\ref{se2} for the final equilibrium of the thermionic charging process.
Across J-I, an electric field builds up as well. It is roughly equal to the
contact potential divided by the width of the depletion region $x_d$, namely the
thin layer across J-I where the diffusion of electrons has taken place.

It is also widely held that as soon as the two halves of the horseshoe
are physically joined at J-I, a potential difference, equal and opposite to the
contact potential $\Delta V_{bi}$ at J-I, should instantaneously build up across
the gap J-II. 

That potential difference must not be confused with that deriving from the
thermionic charging process described in Section~\ref{se2}. It is intended to
be instantaneously generated by the movement of electrons in the bulk of the
horseshoe just after the physical contact of the materials at J-I. People
expect that potential difference across J-II because of an allegedly
straightforward application of Kirchhoff's loop rule to the horseshoe scheme
in Fig.~\ref{fig4}.

The first objection to the TCC functioning is that if this potential
difference really formed across J-II, the thermionic charging process across
plates A and B would not even start. Since $\Delta V$ across J-II is equal
to $(\phi_2 - \phi_1)/e$, no net electron flux from plate B to plate A could be
possible. As described in Section 2, the energy required for an electron to get
extracted from one plate and reach the other plate would be the same (and
incidentally, equal to $\phi_2$). Thus, there would be no net displacement of
electrons across J-II, and therefore there would be no current in the
short-circuit TCC.

My reply to that objection is that there is no instantaneous potential
difference across J-II caused by physically joining the two materials at J-I.
And I can prove it by appealing to the basic definition of potential
difference: the potential difference between points $a$ and $b$ is
defined as minus the work $W$ done by the forces acting upon a test electron
that moves along a(ll) path(s) joining $a$ and $b$, divided by the electron
charge $e$~\cite{dab12a,dab13a,dab13b,dab17}.

\begin{figure}[t]
\begin{center}
\includegraphics[width=8cm]{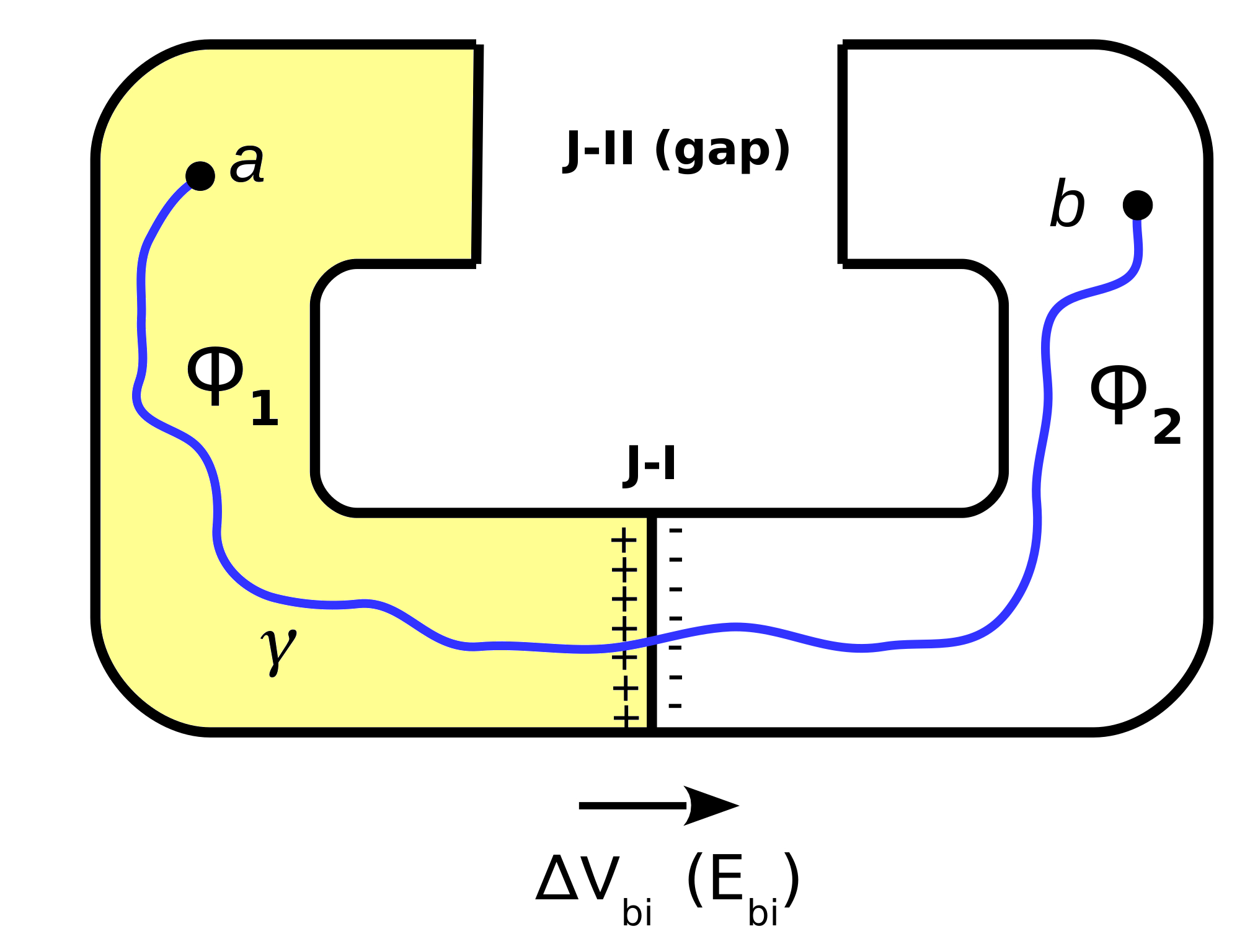}
\end{center}
\caption{Definition of potential difference between points $a$ and $b$ in the
  bulk of the horseshoe}
\label{fig5}
\end{figure}

Let us apply that definition to the previous horseshoe scheme, see
Fig.~\ref{fig5}.  Namely, $\Delta V_{ba} = V_b - V_a =  - W_{ab}/e = -
1/e\int_a^b {\bf F_{int}}\cdot d\pmb{\gamma}$, where ${\bf F_{int}}$ stands for
all the forces acting upon the test electron when it moves along the
path~$\gamma$.

If one considers the path $\gamma$ shown in Fig.~\ref{fig5}, it is evident from
equilibrium considerations that ${\bf F_{int}}$  can be different from zero
only across J-I. When the test electron crosses J-I, it does feel only two
forces: the thermally driven forces ${\bf F_d}$ pushing to the right and the
electric force $e{\bf E_{bi}}$ due to the built-in electric field pushing to
the left, and thus ${\bf F_{int}} = {\bf F_d} + e{\bf E_{bi}}$. By the way, the
test electron does feel the thermally driven forces, just like all electrons
in the depletion region around J-I.

Since the diffusion forces dynamically sustain the electric field at J-I, these
two force fields must always be present across J-I and must always be equal and
opposite at equilibrium, namely ${\bf F_d} = - e{\bf E_{bi}}$. As a matter of
fact, if we switched off the thermally driven forces ${\bf F_d}$, the built-in
electric field ${\bf E_{bi}}$ would go to zero. Therefore, the total force
${\bf F_{int}}$ is zero on average, and so is the potential difference
$\Delta V_{ba}$ between points $a$ and $b$.  

All that means that physically joining two different materials at J-I cannot
instantaneously cause any potential drop across the gap J-II that would kill
the thermionic emission from the outset.

That theoretical result may appear at odds with the fact that a potential
difference $\Delta V$ is actually detected and measured across J-II among
different materials physically joined at J-I. For instance, measuring this
$\Delta V$ is one of the reasons why the Kelvin probe force microscope (KPFM)
has been developed. My explanation for this apparent inconsistency is that KPFM,
in fact, measures a potential difference across J-II, but this potential
difference is just that generated by thermionic emission across J-II. In the
case of a KPFM scan, I am referring to the thermionic emission between the
surface of the scanned material and the Kelvin probe tip. The potential
difference measured by a KPFM is thus not that due to the contact potential
at J-I.

The second objection to the expected functioning of a TCC is that even if a
net flux of emitted electrons were possible from left to right across J-II
(see, Fig.~\ref{fig5}), the current would not flow back across junction J-I.
Junction J-I is known to be a Schottky-type rectifying junction: under biasing,
electrons easily flow from left to right across J-I but not that easily the
other way around (from right to left across J-I).

The reply to that objection is simple and more immediate~\cite{dab11a,dab17}.
It is well known that every Schottky junction is characterized by a
{\em reverse leakage current} that overcomes the rectifying barrier even under
small reverse biasing. That current may be high or low depending upon several
factors. It depends (although slightly) upon the magnitude of the reverse bias.
It also depends upon junction materials and preparation. It is proportional to
the temperature and has an inverse dependence upon the depletion region width
$x_d$. According to the literature, the reverse leakage current may range from
$10^{-8}$~A/cm$^2$ to $10^{-2}$~A/cm$^2$ (there also exist examples of higher
values), and as I will show in the next section, even the lowest value is
already enough to allow the circulation of a current through a short-circuit
TCC (Fig.~\ref{fig3}, right).

\section{Numerical simulation and results}
\label{se4}

\begin{figure}[t]
\begin{center}
  \includegraphics[width=12cm]{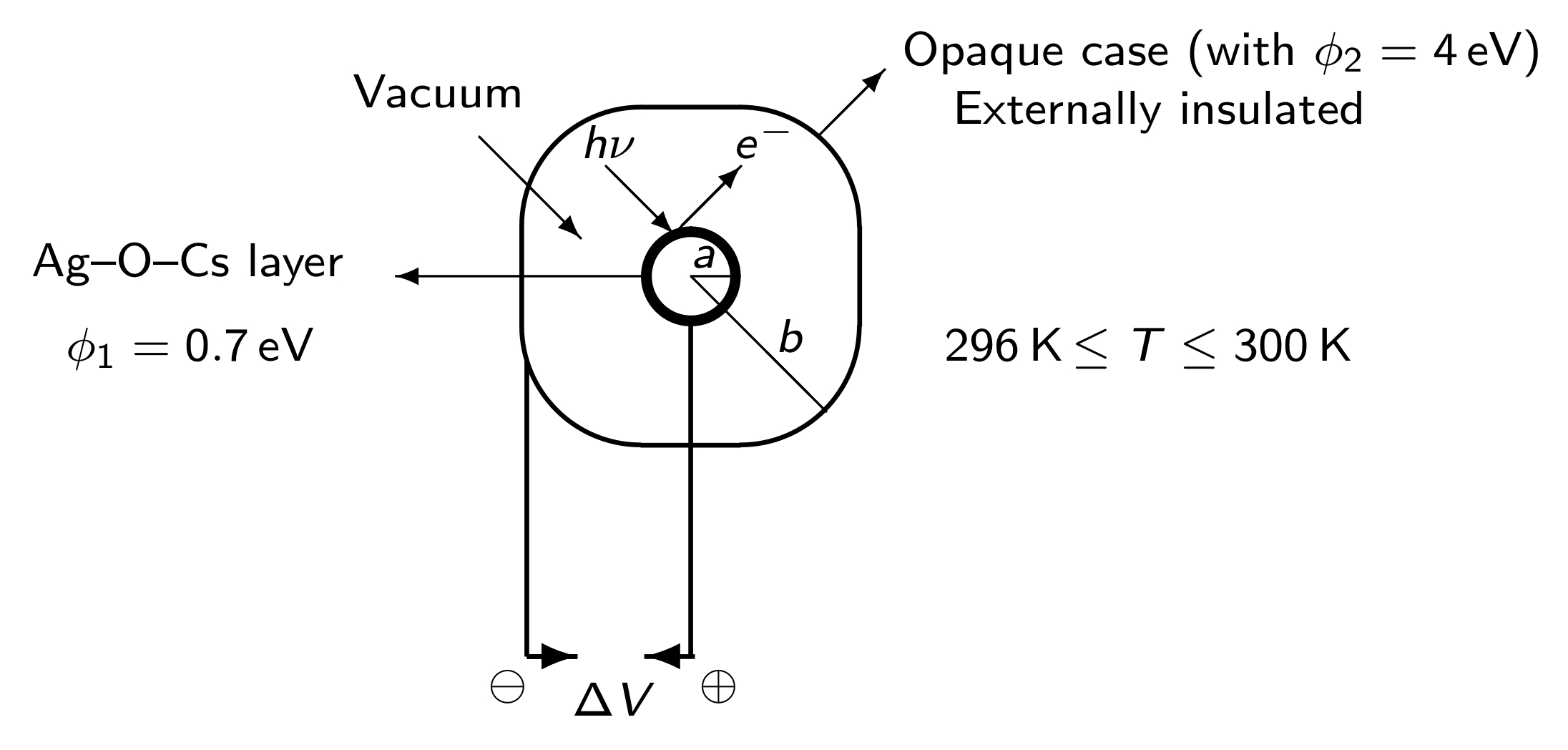}
\end{center}
\caption{Sketch of the modeled open-circuit spherical TCC}
\label{fig6}
\end{figure}

I have mathematically modeled the photoelectric effect induced by the
electromagnetic radiation following Plank's law (blackbody radiation) on both
plates of a spherical capacitor (with different work functions). Here, I
describe the results of the numerical simulation. I have studied a spherical
TCC (see Fig.~\ref{fig6}) because it is easier to treat
analytically~\cite{dab11a,dab17}.

For the low work function coating, I have chosen the well-known photo-cathode
material S1 (Ag-O-Cs) with a work function of nearly 0.7~eV. The inner and
outer spheres are made of metal with a work function of 4~eV. The modeled TCC
is 40~cm wide ($b =20$~cm and $a = 10$~cm). The ratio between the outer and the
inner radius is taken equal to 2 because it can be proven that this choice
maximizes the charging speed in the open-circuit setup~\cite{dab10}.

Another significant parameter for the simulation is the quantum efficiency
$\eta$ of the materials. The quantum efficiency is roughly the ratio between
the number of electrons extracted by the radiation and the number of photons
impinging on the plates. It is a frequency-dependent parameter. I have plugged
into the equations only mean values over frequency. I have used highly
conservative values for the quantum efficiencies to stress-test the robustness
of the process: $\eta_1 = 10^{-5}$ and $\eta_2 = 1$. More realistic values are
$\eta_1 \geq 10^{-3}$ ($\geq 0.1\%$)~\cite{hama} and $\eta_2 \ll 1$. Moreover,
I have found the whole process and results to be quite insensitive to the
exact value of $\eta_2$.

The differential equation that governs the charging process of an open-circuit
spherical TCC is the following~\cite{dab11a,dab17}:

$$ \frac{dV(t)}{dt}= \frac{\pi e b}{2\epsilon_0
c^2}\biggl(\frac{kT}{h}\biggr)^3 \Biggl(\overline{\eta}_{1}\int_{\frac{eV(t) +
\phi_{1}}{kT}}^\infty \frac{x^2 dx}{e^{x}-1} - 
4\overline{\eta}_{2}\int_{\frac{\phi_{2}}{kT}}^\infty
\frac{x^2 dx}{e^x-1}\Biggr), $$

\noindent where $V(t)$ is the voltage between the spheres, $b$ is the radius of
the outer sphere, $c$ is the velocity of light, $e$ is the electron charge,
$\epsilon_0$ is the vacuum permittivity, $h$ is the Planck constant, $T$ is the
absolute temperature, and $k$ is the Boltzmann constant.  

\begin{figure}[t]
\begin{center}
  \includegraphics[width=12cm]{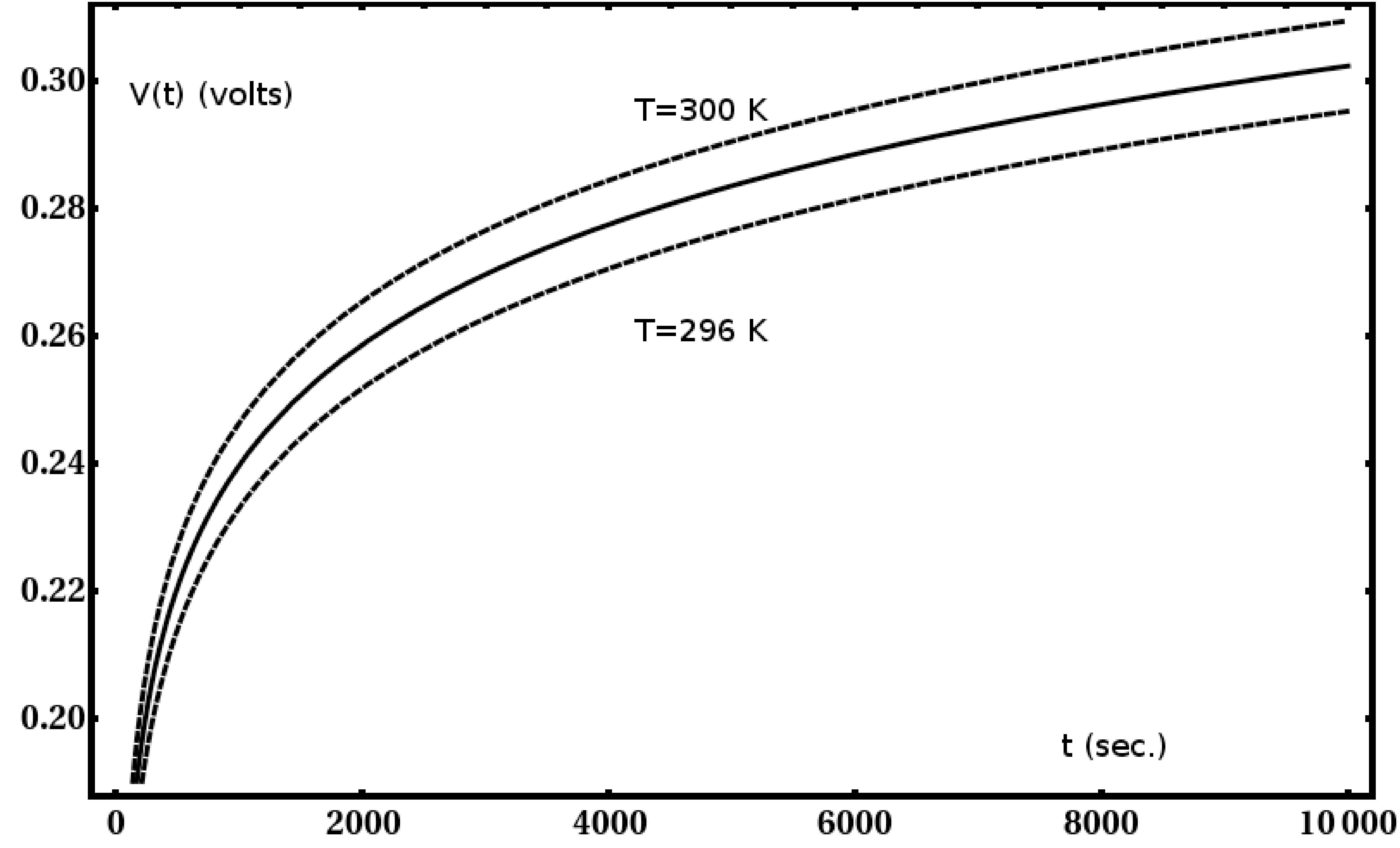}
\end{center}
\caption{Charging profile of the open-circuit spherical TCC described in the
  text. Charging profiles for three absolute temperatures (296~K, 298~K, and
  300~K) are given}
\label{fig7}
\end{figure}

The graph in Fig.~\ref{fig7} shows the charging profile of a spherical
open-circuit TCC with the abovementioned parameters.

With more realistic values for $\eta_1$ and $\eta_2$, the charging process is
expected to be sensibly faster (one or two orders of magnitude faster).

When the terminals of the spherical TCC are connected to a resistor $R$, a
current starts to flow across the circuit and settles to a steady-state value
$i_S$. That current is related to the steady-state voltage $V_S$ across $R$ and
the TCC as follows~\cite{dab11a,dab17}:

$$i_{s}=\frac{2\pi^2 e b^2}{c^2}
\biggl(\frac{kT}{h}\biggr)^3\Biggl(\overline{\eta}_{1}\int_{\frac{eV_s +
\phi_{1}}{kT}}^\infty \frac{x^2 dx}{e^{x}-1} - 4\overline{\eta}_{2}
\int_{\frac{\phi_{2}}{kT}}^\infty \frac{x^2 dx}{e^x-1}
\Biggr),$$

\noindent and, obviously, they satisfy Ohm's law $V_S = R\cdot i_S$.

\begin{figure}[t]
\begin{center}
  \includegraphics[width=12cm]{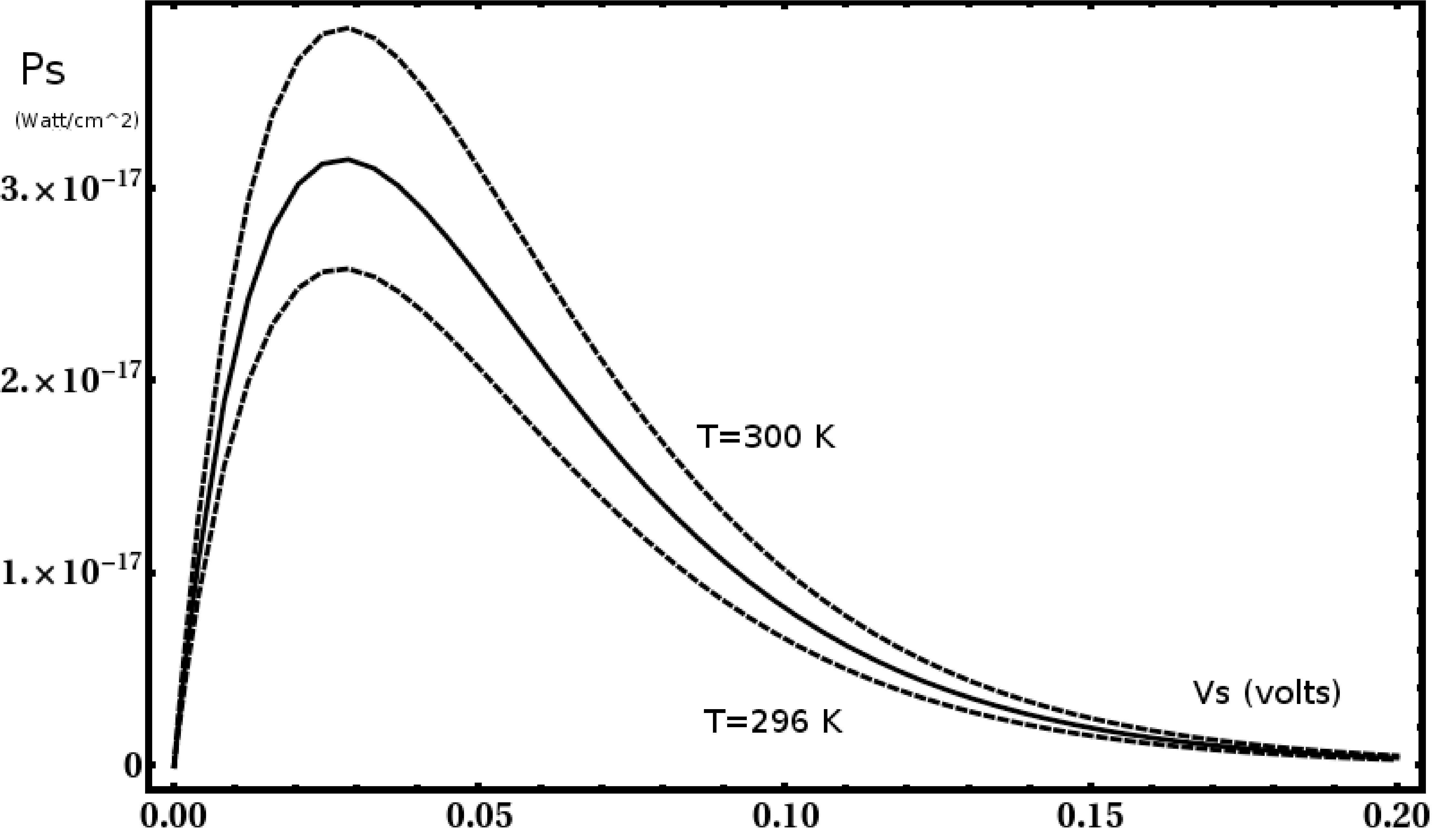}
\end{center}
\caption{Power output per unit area of the inner sphere versus the
  steady-state voltage across the TCC and the resistor $R$ (for three values
  of the ambient temperature)}
\label{fig8}
\end{figure}

The power density $P_S$ of the short-circuit spherical TCC is then given by
$P_S = V_S\cdot i_S / S_a$, where $S_a$ is the surface of the inner sphere.
The graph in Fig.~\ref{fig8} shows the power density $P_S$ as a function of
the steady-state voltage $V_S$ in Watts/cm$^2$.

By tuning the value of the external resistance $R$, we can maximize the power
output. The maximum is obtained at 0.03~V across a resistance of the order of
$10^{10}~\Omega$. Therefore, the maximum power output reachable by the spherical
TCC with the abovementioned physical characteristics is nearly $10^{-14}$~W,
with a steady-state current of the order of $10^{-12}$~A. 

By plugging into the equations more realistic values for $\eta_1$ and $\eta_2$,
I expect to have a power output and a current at least two orders of magnitude
higher, $P_S\approx 10^{-12}$~W and $i_S\approx 10^{-10}$~A.

The power output of the modeled TCC and the power density represented in
Fig.~\ref{fig8} clearly show that the current design has no hope for an
immediate application for large-scale ambient energy harvesting.
However, it is crucial to experimentally prove it since only this way we will
have confirmation that our approach is not at odds with fundamental physics.

\section{About experimental tests}
\label{se5}

Before describing the two experiments I became aware of that seem to implement
the design described above, let me linger over a few necessary (but not
sufficient) prescriptions that every experimental test must comply with to be
credible. 

They are listed below in no particular order:

\begin{itemize}
\item Reduce external disturbance (varying electric and magnetic fields,
  e.m. pickup, cosmic rays, etc.) by properly shielding the TCC and the
  measurement equipment with Faraday cages and mu-metal boxes and by performing
  the test inside an underground laboratory;
\item Carefully control the uniformity of the temperature to prevent
  thermoelectric effects: no significant temperature gradient should be allowed
  in the space hosting the TCC and the measurement equipment;
\item Place two TCCs side by side: they must be the same, except that one does
  not have any emitting layer on its plates;
\item Use an ultra-high input impedance electrometer ($>10^{10}\,\Omega$).
  According to the previous simulation, the resistance external to the TCC must
  be very high. 
\end{itemize}

Another significant check to do would be using multiple equivalent TCCs. They
should be connected in series or parallel to see whether the overall voltage and
current scale up with the number of TCCs. If they do not, then any possible
non-zero voltage and current values for a single TCC could easily be caused
by noise in the measurement equipment. This check would be even more important
than seeing whether the magnitudes of voltage and current on a single TCC
correlate with ambient temperature: even the internal noise of a
high-sensitivity electrometer correlates with ambient temperature.  

\begin{figure}[t]
\begin{center}
  \includegraphics[width=11cm]{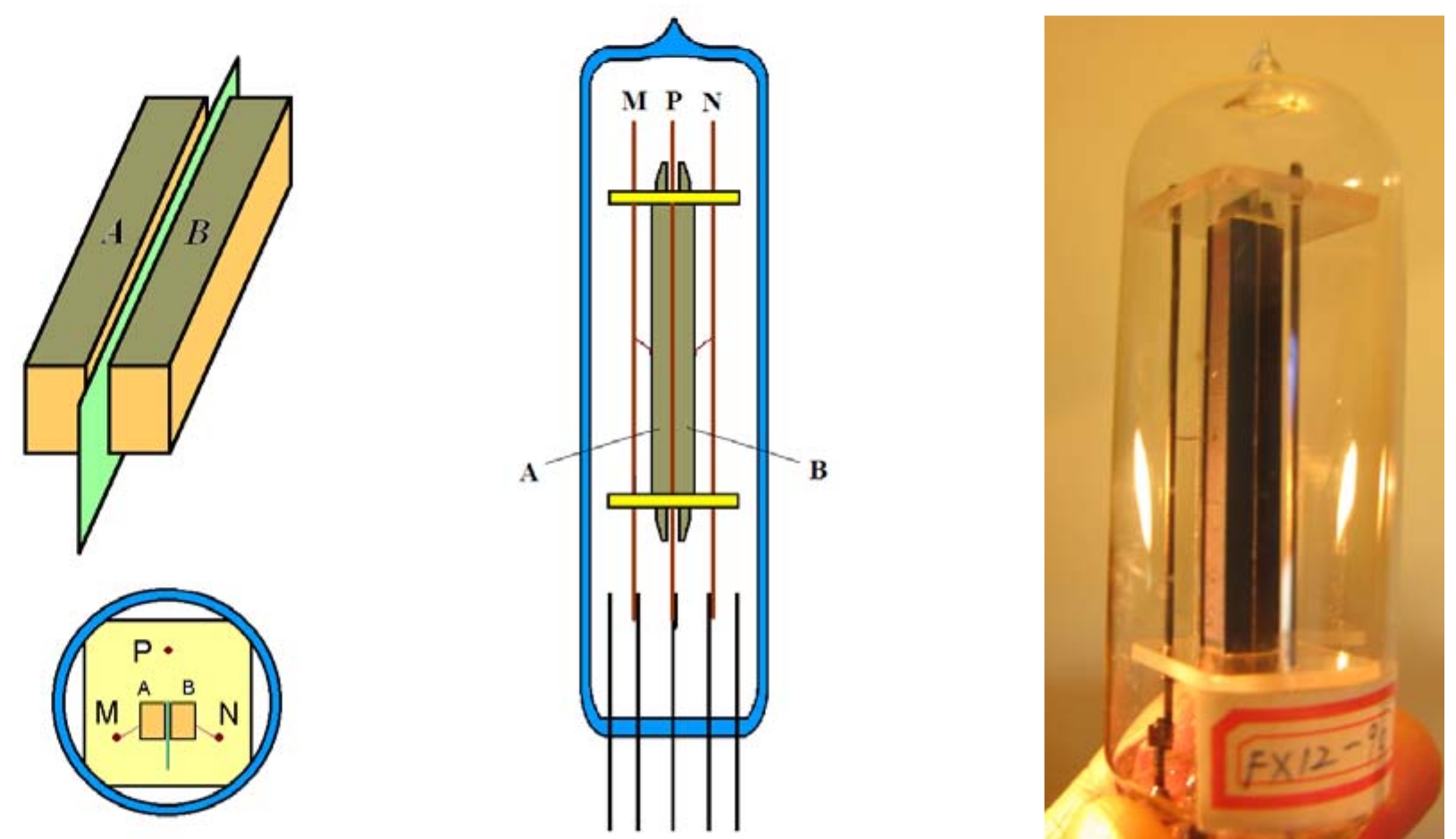}
\end{center}
\caption{One of the vacuum tubes used by Fu and Fu~\cite{fu}}
\label{fig9}
\end{figure}

At the end of 2014, I received, as personal communication, a report by Fu and
Fu~\cite{fu} on an experiment conducted on some Ag-O-Cs vacuum bulbs
with custom design (see Fig.~\ref{fig9}). They were previously employed by the
same authors to test the possibility of drawing energy from ambient
heat with the help of an external static magnetic field~\cite{leff}. In that
report, they actually implemented and tested the basic scheme of a
thermo-charged capacitor like the one sketched in Fig.~\ref{fig1}.

\begin{figure}[t]
\begin{center}
  \includegraphics[width=11cm]{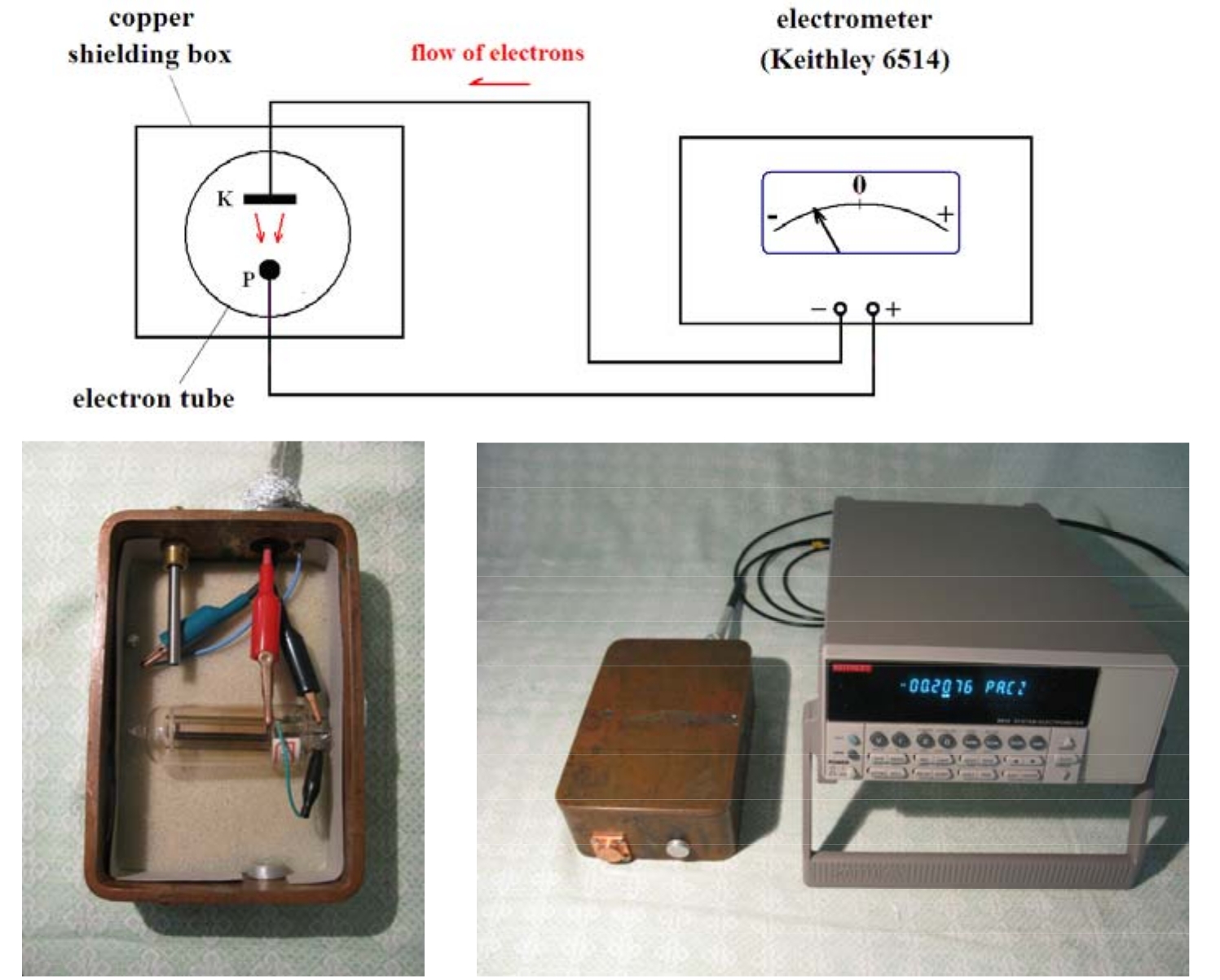}
\end{center}
\caption{Schematic of the experiment carried out by Fu and Fu~\cite{fu}}
\label{fig10}
\end{figure}

How their original vacuum bulbs were manufactured (see Fig.~\ref{fig9})
made Fu and Fu able to use the Ag-O-Cs plates A and B, connected in series, as
the emitter and a cesium-coated molybdenum supporting rod facing the plates (P)
as the collector of a simplified version of a thermo-charged capacitor.
A schematic of the experiment, taken from Fu and Fu's
report, is given in Fig.~\ref{fig10}, where the vacuum tube, a shielding copper
box, and the measuring apparatus (electrometer Keithley 6514) are visible.

Fu and Fu measured a stable current exceeding $2\times 10^{-12}$~A and a
voltage drop of the order of 100~mV. They also verified that both these values
change sign once the electrometer's connections to the tube are switched over.
Unfortunately, they did not investigate the dependence of the current/voltage
output on the number of connected tubes and the value of the external uniform
temperature: the measurements described in the report have been taken on a
single vacuum tube and at a single external temperature. Their results are
extremely interesting and promising and seem to give a direct experimental
corroboration of the theory behind the thermo-charging process. Their vacuum
tube has probably an emitting surface of a few square centimeters. If we scale
down the results presented in the previous Section by two orders of magnitude
in the inner area and use the more realistic value for the quantum efficiency
of the emitting surface ($\eta_1\geq 10^{-3}$), we see that the expected
current for a single vacuum tube is of the same order of magnitude as that
found by Fu and Fu. 

In their report, Fu and Fu also ventured into a theoretical explanation for
their results. However, in my opinion, they did not succeed. Unfortunately,
their explanation is quite sketchy and presents some flaws at a fundamental
level.  

\begin{figure}[t]
\begin{center}
  \includegraphics[width=6cm]{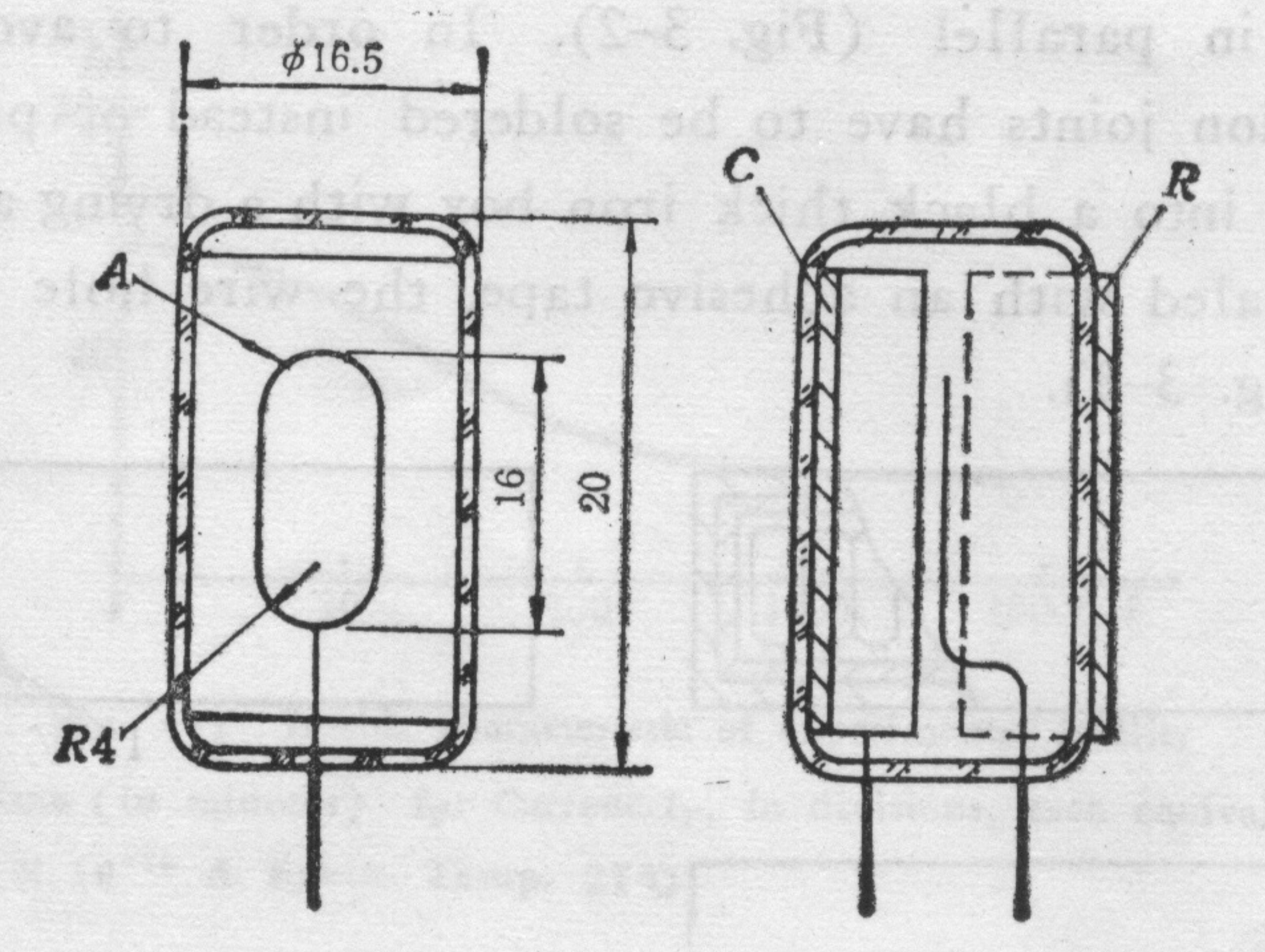}\hspace{1cm}
  \includegraphics[width=6cm]{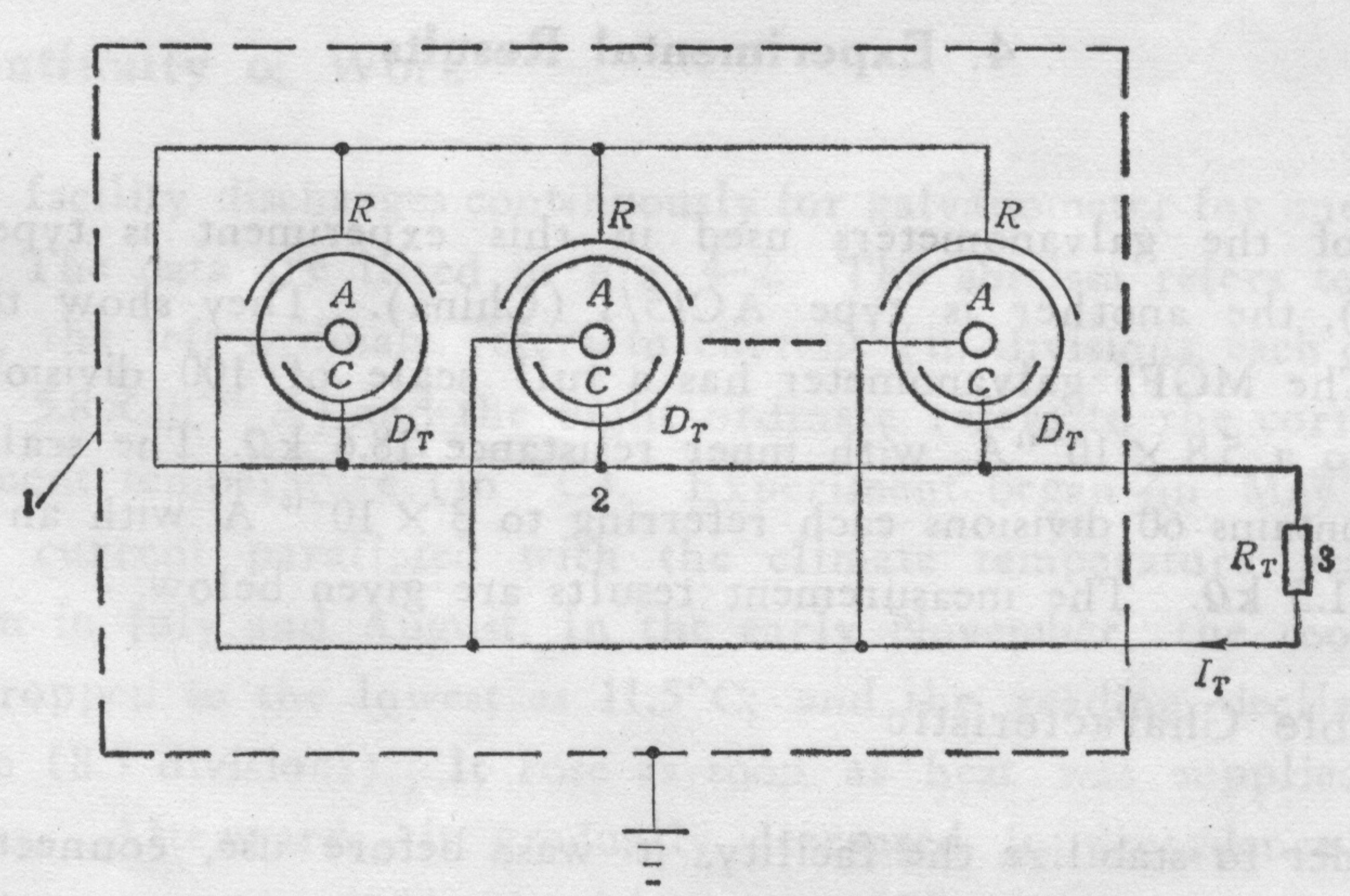}
\end{center}
\caption{Some of the schematics of the experiment carried out by Xu
  Yelin~\cite{yel}}
\label{fig11}
\end{figure}

In May 2011, a person who read my first papers on the TCC sent me an email and
drew my attention to the work of a Chinese researcher, Xu Yelin, which I was
not aware of~\cite{yel}. Yelin performed a large-scale experiment using
nearly 700 custom Ag-O-Cs vacuum tubes connected in series
(see Fig.~\ref{fig11}). Although he did not use any ultra-high input
electrometer, he found non-zero current and voltage  across the circuit. In
particular, at the ambient temperature of 32$^\circ$~C, he measured a peak
short-circuit output current of $\approx 5\times 10^{-8}$~A and a voltage drop
of 55~mV across an external resistance of $10^8\,\Omega$. The interesting fact
is that the current and voltage values remained stable for over a year (the
duration of the experiment), with variations closely matching the ambient
temperature trend. 

Even in this case, the emitting surface of a single vacuum tube is of the order
of a few square centimeters, and thus the current generated by each tube is of
the order of $10^{-12}$~A. Again, it seems to agree with the results of the
simulations presented in the previous Section with $\eta_1\geq 10^{-3}$. 

Yelin proposed a theoretical explanation for his results, but again it appears
to be somewhat cursory and plagued by several flaws at a fundamental level.
That notwithstanding, Yelin’s experiment and results remain fascinating.  

To conclude this section, let me describe an indirect and easier way to test the
expected functioning of a TCC~\cite{dab17}. Let us use the TCC as a
conventional photoelectric device and exclusively  illuminate the low work
function plate (plate B in Fig.~\ref{fig1}) with direct light of arbitrary
intensity but with frequency $\nu < \phi_2/h$. The fact that plate A is not
illuminated does not count since the light used has a too low frequency to
extract electrons from it. In fact, that would be almost like having a
TCC immersed in blackbody radiation with a higher intensity than usual
and a frequency cutoff at $\phi_2/h$.

It should be easy to understand that if we can measure a non-zero potential
difference and a non-zero current in the open-circuit and short-circuit
configurations, respectively, then the TCC should work as expected even when it
is immersed in uniform blackbody radiation as in the original design. The only
difference would be in the rapidity with which the potential difference is
generated or the magnitude of the current in the short-circuit configuration.
Those would be more favorable in the case of the TCC used like a photoelectric
device. All this would also help with the measuring procedure.
By suitably increasing the intensity of the impinging direct light, the current
flowing across a short-circuit TCC can be made macroscopic, and an ultra-high
input impedance electrometer would not be needed.

\section{Concluding remarks}
\label{se6}

From a fundamental physics perspective, the theory behind the photoelectric
effect induced by blackbody radiation is pretty solid. Moreover, at least two
experimental tests have been performed in the past that seem to corroborate,
although not definitely, the alleged functioning of the proposed process. More
dedicated and specifically designed experiments are needed to get  definitive
confirmation, along with the simple, although indirect, test described at the
end of the previous section. However, due to the extremely low power density
output, even in the case of positive experimental results, the question of the
immediate application for ambient energy harvesting remains open.  

\section*{Acknowledgments}
I wish to thank Prof.~Daniel Sheehan and Prof.~Garret Moddel for the invitation
to the inspiring SSE Symposium on Advanced Energy Concepts Challenging the
Second Law of Thermodynamics held in January 2022 (symposium hosted as part of
the $4^{th}$ Annual Advanced Propulsion and Energy Workshop). The present paper
summarizes the presentation given at that symposium. I am also grateful
to Dr.~Gianpietro Summa for his comments on an early draft of the manuscript.


\end{document}